# Exploring User Acceptance of Blockchain-Based Student Certificate Sharing System: A Study on Non-Fungible Token (NFT) Utilization

Prakhyat Khati[1[0009-0000-1383-888X]], Ajay Kumar Shrestha[2[0000-0002-1081-7036]], Julita Vassileva[1[0000-0001-5050-3106]]

[1] University of Saskatchewan, Saskatoon, SK S7N 5C9, Canada
`prakhyat.khati@usask.ca, jiv@cs.usask.ca`
[2] Vancouver Island University, Nanaimo, BC V9R 5S5, Canada
`ajay.shrestha@viu.ca`

**Abstract.**
Blockchain technology has emerged as a transformative tool for data management in a variety of industries, including fintech, research and healthcare. We have developed a workable blockchain-based system that utilizes non-fungible tokens (NFTs) to tokenize and prove ownership of the academic institution's credentials. This makes it easier to create provenance and ownership documentation for academic data and meta credentials. This system enables the secure sharing of academic information while maintaining control, offering incentives for collaboration, and granting users full transparency and control over data access. While the initial adoption of these systems is crucial for ongoing service usage, the exploration of the user acceptance behavioral model remains limited in the existing literature. In this paper, we build upon the Technology Acceptance Model (TAM), incorporating additional elements to scrutinize the impact of perceived ease of use, perceived usability, and attitude towards the system on the intention to use a blockchain-based academic data and meta credentials sharing system. The research, grounded in user evaluations of a prototype, employs a TAM-validated questionnaire. Results indicate that individual constructs notably affect the intention to use the system, and their collective impact is statistically significant. Specifically, perceived ease of use is the sole factor with an insignificant influence on the intention to use. The paper underscores the dominant influence of attitude towards the system on perceived usefulness. It concludes with a discussion on the implications of these findings within the context of blockchain-based academic data and meta credentials sharing, incorporating NFTs for ownership definition.

# 1  Introduction

In recent years, the adoption of blockchain technology has garnered significant attention across diverse sectors [1]. One domain where blockchain's potential is poised for transformation is in educational data sharing, specifically concerning academic certificates and student transcript data [1].Representing student academic datasets as Non-Fungible Tokens (NFTs) on a blockchain ensures the uniqueness and immutability of the datasets, thereby certifying their authenticity and safeguarding against tampering or fraudulent alterations. The introduction of NFTs into educational certificate management represents an emerging frontier. As more students travel overseas to further their education or look for work, a safe and effective mechanism for exchanging academic credentials is essential. Traditional paper-based certification processes are time-consuming and costly, with multiple steps for certification, translation, and authentication. Existing web 2.0-based centralized systems for certificate sharing face challenges such as a lack of global standardization and limited accessibility [1]. The exploration of blockchain-based solutions, particularly the use of NFTs, aims to address these challenges.

This paper presents a comprehensive evaluation of the blockchain-based prototype model which we implemented [2]. It employs the Technology Acceptance Model (TAM) to assess user attitudes toward the proposed system. TAM evaluates factors such as perceived ease of use, perceived usefulness, attitude toward the system, and intention to use. The methodology, research questions, and hypotheses are outlined, providing insights into the experimental design and participant demographics. Applying TAM, the study investigates user acceptance and trust in the blockchain-based application. The subsequent sections detail the findings, emphasizing the significant direct and indirect effects of TAM constructs on the intention to use the system. In conclusion, the paper underscores the pivotal role of TAM in understanding user attitudes toward blockchain technology, particularly NFTs, in reshaping academic data sharing and certificate management.

Our study examines how perceived ease of use (PEOU), perceived usefulness (PU), and attitude towards the system (ATS) relate to users' intention to use (ITU) a blockchain-based student certificate-sharing system. We found direct influences among these factors, with ATS significantly affecting both ITU and PU. PEOU significantly impacts ATS and PU, while its direct effect on ITU is not statistically significant. PU also significantly impacts ITU.

Moreover, our analysis delves into indirect paths, revealing additional insights. PEOU indirectly affects PU through ATS and influences ITU through PU. Additionally, there's a significant indirect effect where PEOU shapes ITU through ATS. Interestingly, the indirect effect of ATS on ITU through PU is not statistically significant, indicating a direct influence. Similarly, the indirect effect of PEOU on PU, subsequently affecting ITU through ATS, is not statistically significant.

In addition, the analysis underscores the collective influence of Perceived Ease of Use (PEOU), Perceived Usefulness (PU), and Attitude Towards System (ATS) on Intention to Use (ITU). We review blockchain-based certificate systems and extend the Technology Acceptance Model (TAM). Section III outlines our methodology, including surveys. Section IV covers the analysis of the results. Conclusions and future research are discussed in Section V.

## 2     Background and Related Works

Blockchain technology has the potential to transform certificate issuance and verification. Early platforms allowed universities to issue and back up certificates but did not utilize NFTs [2]. Later, NFTCert was proposed as a framework to address challenges in existing blockchain-based education certificate-sharing systems [3]. The development of using NFTs for managing digital arts prompted exploration into their potential benefits for certificates and badges [4]. While NFT-based systems like NFTCert and "Ethernal Digital Certificates" have reduced fraudulent activities, they have limitations. For instance, NFTCert restricts the minting of NFTs to institutes, potentially causing time-consuming processes for students and might lead to their dependency on the institute [3]. Additionally, there's ambiguity regarding ownership transfer to student addresses, potentially resulting in challenges or uncertainties regarding the rightful ownership of certificates. Moreover, students cannot create purpose-centric views of certificates, limiting their relevance in specific job applications. NFTCert focuses on NFT certificate management rather than sharing, and its private network implementation may not meet the needs of all stakeholders [3].

In contrast to these approaches, in our previous work [2], we developed a practical and user-friendly blockchain-based model for sharing student credentials and certificates in the educational domain, addressing the challenges. A public blockchain was used to provide accountability for access control, sharing certificates, maintaining complete and updated information, and a verifiable record of provenance, including all accesses/sharing/usages of the data. Students have control over their credentials even after sharing them. Also, after getting their NFT credentials, students can mint more certificate NFTs, i.e., ViewNFTs, where students can tailor their certificate and credentials as per their requirements. Additionally, students are given more control over their credentials, making the platform student focused.

The TAM, derived from the Theory of Reasoned Action (TRA) by Ajzen and Fishbein [5], is widely used to understand users' adoption of information systems across various domains like health, business, and education. However, despite previous research applying TAM to blockchain and smart contract applications [6], there's still a notable gap in understanding its application to Non-Fungible Tokens (NFTs). NFTs, distinct digital assets without direct exchangeability like cryptocurrencies, have gained attention, particularly in the art and collectibles market, due to their ability to authenticate ownership via blockchain technology. Within TAM, it's crucial to note that a system's actual usage (AU) depends on the user's intention to use (ITU), with factors like perceived usefulness and ease of use influencing this intention. Hence, investigating the factors influencing users' acceptance and adoption of NFTs within TAM would help bridge this gap and understand NFT adoption dynamics, with researchers extending TAM by including additional constructs such as Perceived Privacy, Perceived Security, Quality of the System, Attitude towards the system and Trust [6, 7].

TAM suggests that attitude and intention to use an application are determined by assessing the system's perceived usefulness against its perceived difficulty. Perceived usefulness reflects how much a technology enhances performance, while perceived ease of use indicates the effortlessness of using a specific system. TAM-based frameworks are widely used to gauge user acceptance for applications, with validated surveys comprising items related to TAM constructs serving as research instruments. Based on insights from earlier studies by Davis and colleagues [8, 9], similar research initiatives have been undertaken, resulting in extended TAM models. This study explores user attitudes toward acceptance of blockchain-based applications, with similar studies like the one by Shrestha et al. [6, 7] providing valuable insights.

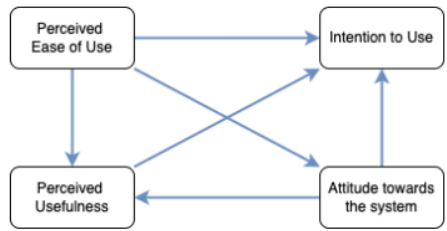

**Fig. 1.** An Extended TAM model for our study.

## 3   Methodology

Our research focused on assessing user feedback for the prototype of a blockchain-based certificate-sharing system, using a questionnaire as a research tool to gather data. We formulated several hypotheses based on a literature review, targeting specific constructs outlined in the table below. The survey instrument consisted of questions, referred to as "items," related to the Technology Acceptance Model (TAM) constructs. These included perceived ease of use (6 items), perceived usefulness (5 items), intention to use (2 items) and attitude towards the system (2 items). Participants expressed their opinions on a 7-point Likert scale, ranging from strongly disagree (1) to strongly agree (7). The detailed list of items corresponding to each TAM construct can be found in Table 1. In this paper, the following hypotheses are being investigated:

- H1: Perceived ease of use significantly influences the perceived usefulness of the system.
- H2: Attitude towards the system significantly influences perceived usefulness.
- H3: Perceived ease of use significantly influences intention to use the system.
- H4: Perceived usefulness significantly influences intention to use the system.
- H5: Attitude towards the system significantly influences intention to use.
- H6: Perceived ease of use significantly influences attitude to use the system.

- H7: The combined effect of perceived ease of use, perceived usefulness, and attitude towards the system significantly influences intention to use.

### 3.1 Experiment Design

We conducted an experiment to assess user acceptance and usage of our blockchain-based prototype model for academic data sharing. Participants used a fully deployed prototype of the NFT-based certificate-sharing system. To understand the likelihood of user adoption, we investigated the influence of perceived usefulness and perceived ease of use on users' attitudes and intentions. Participants were recruited through an advertisement on the university portal "PAWS," targeting current students. Eligible and interested participants provided consent and engaged in the study within a controlled computer lab setting.

**Table 1.** Construct and Items

| Construct & Definition | Items |
|---|---|
| Perceived Ease of Use (**PEOU**) "Degree to which one believes that using a certain system would be free of effort." | **peou1:** Learning to operate this system is easy. <br> **peou2:** I find it easy to get this system to do what I want it to do. <br> **peou3:** My interaction with this system is clear and understandable. <br> **peou4:** I find this system to be flexible to interact with. <br> **peou5:** I feel it is easy to become skillful at using this system. <br> **peou6:** I find this system easy to use. |
| Perceived Usefulness (**PU**) "Degree to which a person believes that using a particular system would enhance his or her job performance." | **pu1:** Using this system would improve performance in certificate sharing with transparency and privacy. <br> **pu2:** Using this system would increase effectiveness in privacy policy formulation. <br> **pu3:** Using this system would make it easier for me to set certificate sharing preferences. <br> **pu4:** Using this system would increase productivity in certificate sharing with more control over privacy. <br> **pu5:** I find this system useful for setting my certificate sharing preferences. |
| Attitude towards the system (**ATS**) "Degree to which one has favorable/unfavorable evaluation of a system." | **ats1:** I believe that using the blockchain-based system would be beneficial for me. <br> **ats2:** In my opinion, it would be desirable for me to use the blockchain-based system. <br> **ats3:** It would be good for me to use the blockchain-based system |
| Intention to Use (**ITU**) "Degree to which a person has a behavioral intention to adopt the technology" | **itu1:** I would like to use this system to set certificate sharing preferences. <br> **itu2:** I would enjoy using this system when I need to use it. <br> **itu3:** It is worthwhile to use this system to set certificate sharing preferences. <br> **itu4:** I will use this system to decide how my data is shared. |

Before the survey, participants received a brief overview video on blockchain technology and NFTs. They interacted with the system actively, signing up with a wallet and exploring different user roles (University, Student, Employer) through a second explanatory video. Participants then assessed the system's usability and effectiveness by creating certificates, filtering subjects, and sharing them with other institutions. The survey, administered via SurveyMonkey, gauged participants' attitudes and intentions related to the prototype model.

Following the interaction, participants completed the survey questions (Table 1), providing quantitative data on perceived usefulness, perceived ease of use, attitude towards the system, and intention to use. Qualitative feedback was also collected to identify specific challenges and opportunities for improvement. A pilot study involving five quantitative research experts from the University of Saskatchewan helped refine the experiment design and survey questionnaire. Ethical approval for the research protocol was obtained from the University of Saskatchewan's Behavioral Research Ethics Board. A total of 40 participants took part in the study, primarily students with diverse levels of familiarity with blockchain technology, smart contracts, and NFTs. Around 33.3% of the participants were somewhat familiar with blockchain and smart contract technology, and 63% were very familiar with them. Similarly, 28.21% were somewhat familiar with the concept of NFT, and 62% were highly familiar with it, but not in the context of applications related to academic certificates.

## 4     Result

This section presents and briefly analyzes the collected data using descriptive statistics. Subsequently, we delve into the results of the structural equation model (SEM), covering both the measurement model and structural models. Data analysis was conducted using Microsoft Excel and SmartPLS software. Respondent scores, measured on a 7-level Likert scale, were categorized for analysis purposes, with ranges including Extremely Low ($0 < x \leq 1$), Quite Low ($1 < x \leq 2$), Slightly Low ($2 < x \leq 3$), Neither ($3 < x \leq 4$), Slightly High ($4 < x \leq 5$), Quite High ($5 < x \leq 6$), and Extremely High ($6 < x \leq 7$). As depicted in Fig. 2, the average results consistently fall within the range of 5 to 6, positioning them within the Quite High category. Tables 2 to 5 summarize the data for each TAM construct: perceived ease of use, perceived usefulness, attitude towards the system, and intention to use.

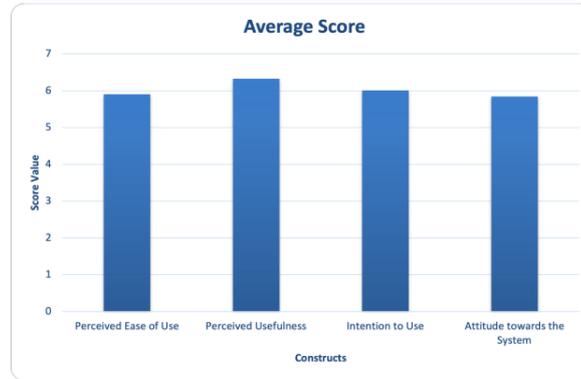

**Fig. 2.** Analysis of the construct

### 4.1 Measurement models

We conducted an exploratory factor analysis to evaluate the measurement model's internal consistency, reliability, and construct validity, focusing on the relationship between observed variables (items) and latent constructs. Strong factor loadings (>0.50) were found for all constructs, including "Attitude towards the System" and "Intention to Use," indicating robust associations. Similarly, "Perceived Ease of Use" and "Perceived Usefulness" also demonstrated strong factor loadings, supporting construct validity. Reliability was confirmed by high Cronbach's Alpha values for each construct: "Attitude towards the System" (0.936), "Intention to Use" (0.937), "Perceived Ease of Use" (0.915), and "Perceived Usefulness" (0.885). Additionally, Composite Reliability (CR) values were notably high for all constructs: "Attitude towards the System" (0.959), "Intention to Use" (0.955), "Perceived Ease of Use" (0.934), and "Perceived Usefulness" (0.917). Convergent validity was assessed through Average Variance Extracted (AVE) values, which exceeded the recommended threshold of 0.50 for all constructs: "Attitude towards the System" (0.886), "Intention to Use" (0.842), "Perceived Ease of Use" (0.704), and "Perceived Usefulness" (0.690). These findings validate the measurement model's effectiveness in assessing users' attitudes, intentions, and perceptions regarding system acceptance.

**Table 2.** Reliability and Validity Measures

| Construct | α | CR | AVE |
|---|---|---|---|
| Attitude towards the System | 0.936 | 0.959 | 0.886 |
| Intention to Use | 0.937 | 0.955 | 0.842 |
| Perceived Ease of Use | 0.915 | 0.934 | 0.704 |
| Perceived Usefulness | 0.885 | 0.917 | 0.690 |

## 4.2 Structural Model

We developed a structural model featuring four key constructs—perceived ease of use, perceived usefulness, attitude toward the system, and intention to use. This model, depicted in Figure 3, explores their potential impact on user adoption of a proposed certificate-sharing framework utilizing NFT and blockchain. Structural Equation Modeling (SEM) analysis was conducted to assess the strength of relationships, calculating coefficients of determination ($R^2$) and path coefficients ($\beta$). These coefficients signify the direction and strength of relationships between constructs. We followed Chin's guidelines [10], where a path coefficient $\geq 0.2$ is considered relevant. A model is deemed statistically significant if the p-value is $< 0.5$, quite significant if $p < 0.01$, and highly significant if $p < 0.001$[11]. The direct path coefficient analysis (see Table 3) revealed that the attitude towards the system (ATS) significantly influences the intention to use (ITU) ($\beta = 0.374$, $p = 0.023$) and the perceived usefulness (PU) ($\beta = 0.457$, $p = 0.006$). Additionally, perceived usefulness (PU) significantly influences the intention to use (ITU) ($\beta = 0.452$, $p = 0.03$).

While perceived ease of use (PEOU) significantly influences perceived usefulness (PU) ($\beta = 0.478$, $p < 0.001$), its direct effect on ITU was not statistically significant ($\beta = 0.156$, $p = 0.165$). However, PEOU significantly affects ATS ($\beta = 0.668$, $p < 0.001$).

In the indirect path coefficient analysis (see Table 4), several significant findings emerged. PEOU significantly influences PU through its impact on ATS ($\beta = 0.305$, $p = 0.012$). Additionally, PEOU indirectly affects ITU through PU ($\beta = 0.216$, $p = 0.036$) and influences ITU through ATS ($\beta = 0.250$, $p = 0.032$). However, the indirect effect of ATS on ITU through PU is not statistically significant ($\beta = 0.206$, $p = 0.094$). Similarly, the indirect effect of PEOU on PU, subsequently affecting ITU through ATS, is also not statistically significant ($\beta = 0.138$, $p = 0.110$).

The total effect path coefficient analysis (see Table 5) combines both direct and indirect effects. The combined effects of PEOU, PU, and ATS on ITU were statistically significant ($p < 0.05$), with path coefficients ranging from 0.452 to 0.784 and ($R^2 = 0.821$) supporting hypothesis H7.

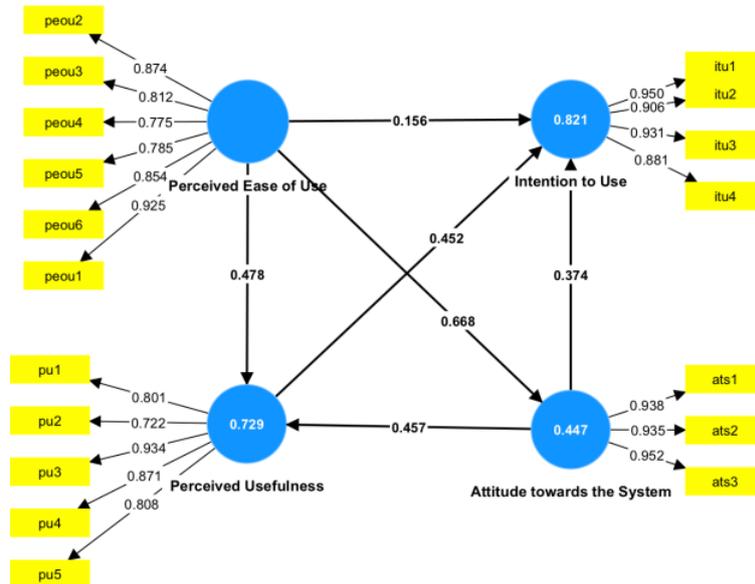

**Fig. 3.** Structural Equation Modeling (SEM) Analysis of all Constructs

**Table 3.** Direct Path Coefficient Analysis

|  | Path Coefficient | Standard Dev | P Value |
|---|---|---|---|
| ATS ➡ ITU | 0.374 | 0.187 | 0.023 |
| ATS ➡ PU | 0.457 | 0.180 | 0.006 |
| PEOU ➡ ATS | 0.668 | 0.133 | 0.000 |
| PEOU ➡ ITU | 0.156 | 0.161 | 0.165 |
| PEOU ➡ PU | 0.478 | 0.142 | 0.000 |
| PU ➡ ITU | 0.452 | 0.241 | 0.030 |

**Table 4.** Indirect path coefficient analysis

|  | Path Coefficient | Standard Dev | P Value |
|---|---|---|---|
| PEOU ➡ ATS ➡ PU | 0.305 | 0.135 | 0.012 |
| PEOU ➡ PU ➡ ITU | 0.216 | 0.122 | 0.036 |
| PEOU ➡ ATS ➡ ITU | 0.250 | 0.135 | 0.032 |
| ATS ➡ PU ➡ ITU | 0.206 | 0.157 | **0.094** |
| PEOU ➡ ATS ➡ PU ➡ ITU | 0.138 | 0.113 | **0.110** |

**Table 5.** Total effect path coefficient analysis

|  | Path Coefficient | Standard Dev | P Value | $R^2$ |
|---|---|---|---|---|
| ATS ➡ ITU | 0.581 | 0.149 | 0.000 | 0.821 |
| PU ➡ ITU | 0.452 | 0.241 | 0.030 |  |
| PEOU ➡ ITU | 0.761 | 0.108 | 0.000 | |
| ATS ➡ PU | 0.457 | 0.180 | 0.006 | 0.729 |
| PEOU ➡ PU | 0.784 | 0.0.98 | 0.000 | |
| PEOU ➡ ATS | 0.668 | 0.133 | 0.000 | 0.447 |

## 5   Discussion

While it's not always feasible to explore every system in detail within one paper, this research aims to offer a broadly applicable framework, providing valuable insights into technology acceptance and usability within credential systems. Overall, our research assessed user experience with a system for certificate sharing via Non-Fungible Tokens (NFTs) and blockchain technology, using the Technology Acceptance Model (TAM) and incorporating an external construct into the conventional TAM framework. Hypotheses H1, H2, H4, H5, H6, and H7 were supported based on significant path coefficients and p-values, revealing a significant link between user attitude, perceived usefulness, and intention to use the system. Hypothesis H3 was not supported, as the direct path coefficient for PEOU on ITU was not statistically significant ($\beta = 0.156$, $p = 0.165$). This can be explained by the prototype's limited functionality. Challenges in evaluating such platforms include resource-intensive development, with utilizing a prototype for partial evaluation emerging as the most practical approach. Our study's reliance on a small sample size, mainly comprising students less familiar with blockchain concepts, presents another limitation. Adoption challenges for distributed ledger-based systems, like NFT-based certificate sharing, include skill gaps, limited organizational awareness, and distrust in blockchain security.

## 6   Conclusion

We introduce an extended TAM-based model to assess user acceptance of a prototype certificate-sharing system utilizing NFT and blockchain technology. Through descriptive statistics, measurement, and structural models, including SEM analysis, we identify predictors of user acceptance. Our work is one of the first to employ TAM to examine user acceptability in the context of a blockchain-based NFT academic certificate system. By employing theory-based model building, user studies, and statistical analysis, we uncover factors influencing intention to use and platform adoption, paving the way for further exploration of NFT and blockchain applications through user behavior modeling. This study demonstrates our use of theory-informed modeling and simulated use case studies to evaluate the adoption potential of complex technologies.

**Acknowledgement**
This research was funded by the NSERC Discovery Grant RGPIN RGPIN-2021-03521 of the third author.

**Disclosure of Interests**

The authors have no competing interests to declare that are relevant to the content of this article.